\documentclass{article}
\usepackage{amsmath}
\usepackage{amsfonts}
\usepackage{graphicx}
\usepackage{color}
\usepackage{subfigure}
\begin{document}
\title{Fokker-Planck Equation of Schramm-Loewner Evolution}
\author{M. N. Najafi}
\author{M. N. Najafi\\
\footnotesize{Department of Physics, University of Mohaghegh Ardabili, P.O. Box 179, Ardabil, Iran}}
\date{}
\maketitle
\begin{abstract}
In this paper we statistically analyze the Fokker-Planck (FP) equation of Schramm-Loewner evolution (SLE) and its variant SLE($\kappa,\rho_c$). After exploring the derivation and the properties of the Langevin equation of the tip of the SLE trace, we obtain the long and short time behaviors of the chordal SLE traces. We analyze the solutions of the FP and the corresponding Langevin equations and connect it to the conformal field theory (CFT) and present some exact results. We find the perturbative FP equation of the SLE($\kappa,\rho_c$) traces and show that it is related to the higher order correlation functions. Using the Langevin equation we find the long-time behaviors in this case. The CFT correspondence of this case is established and some exact results are presented.
\end{abstract}

\section{Introduction}

Schramm-Loewner evolution (SLE) is a stochastic process which aims to classify the two-dimensional critical models by focusing on their geometrical quantities \cite{Shramm,cardy}. The idea of classification of probability measures of random curves growing in the upper half complex plane was first proposed by O. Schramm \cite{Shramm}. This technique has provided us with a new interpretation of the traditional conformal field theory (CFT) and Coulomb gas approaches. In Conformal field theories (CFT) the emphasis is on the algebraic properties of the critical models dictated by their conformal invariance \cite{Dif}. Therefore it seems reasonable that there is a connection between this theory and SLE. This relation was found in the pioneering work of M. Bauer and D. Bernard \cite{BauBer} in which it was found a relation between the central charge of CFT and the diffusivity parameter ($\kappa$) of SLE.\\
Many papers have been published concerning the analytical and numerical properties of interfaces of various two-dimensional (2D) critical models described by SLE theory \cite{Shramm,BerBofCelFal,BerBau,NajafiSAW}. The statistical analysis of the various aspects of these critical curves contains valuable information of the model in hand. In the numerical studies of the less-known 2D critical models, the main concern is to extract the diffusivity parameter (referred to as $\kappa$) of their critical interfaces to relate the underlying model to a CFT class \cite{BerBofCelFal,NajafiSLEkr}. Therefore the dependence of statistical observables to $\kappa$ is of paramount importance in this area \cite{cardy,NajafiLPP}. As an example of these statistical observables, one can mention Fokker-Planck (FP) equation \cite{Gruzberg}, left passage probability \cite{NajafiLPP}, crossing probability \cite{cardy}, etc. which are directly related to the diffusivity parameter ($\kappa$) of SLE. In spite of many analysis addressing the FP equation for the SLE traces in various papers \cite{ViklundLawler,Beffara}, a complete and comprehensive analysis of this subject is missing in the literature yet, e.g. the actual behavior of the solutions of this equation in the whole domain of the model and a comprehensive investigation of the FP equation for the dipolar and radial SLE and SLE($\kappa,\rho_c$) \footnote{It is customary to use the phrase SLE($\kappa,\rho$). In this paper we use SLE($\kappa,\rho_c$) in order not to be confused with the probability density $\rho(x,y,t)$ to be defined in SEC \ref{FP}. Our analysis is limited to the case $\rho_c=\kappa-6$.}. The question \textit{what the statistical properties of the tip of the growing random curves are in the SLE($\kappa$) and SLE($\kappa,\rho_c$)} is the main concern of this paper. The importance of the statistical analysis (such as FP equation) of SLE($\kappa,\rho_c$) can be understood, e.g. for the case in which the curve in its growth process \textit{experiences} some boundary condition changes (bcc) in some point on the boundary of the domain of growth. This situation which occurs in most cases in the numerical analysis of the less-known models should be investigated within SLE($\kappa,\rho_c$) theory. Such analysis may help to analyze these behaviors. \\
In this paper we analyze the various aspects of the FP equation of SLE traces. To this end, we firstly show how to employ the backward SLE equation and find the Langevin equation for the spatial coordinates of the tip of the SLE trace. We analyze these equations in short and long time limits and show that for ordinary (chordal) SLE in long time both horizontal and vertical components of the tip of the SLE trace behave as $t^{\frac{1}{2}}$ ($t=$ time) as expected. To do so, we find the behaviors of SLE traces for long times in two ways; directly solving the FP equation and using the Langevin equation. We then connect the FP equation to the CFT and find some exact behaviors. For SLE($\kappa,\rho_c$), however there is other preferred point on the real axis, causing the situation be different. To analyze this case we first find the behavior of the curves for short times perturbatively. The long-time behaviors are also investigated by analyzing the Langevin equation. We finally report some exact behaviors of this case by means of its CFT counterpart.\\
The paper has been organized as follows: in SEC \ref{SLE} we present a short introduction to SLE($\kappa$) and its variant SLE($\kappa,\rho_c$). The SEC \ref{FP} has been devoted to the derivation of FP equations for the SLE traces. FP equation for the SLE($\kappa,\rho_c$) is derived in SEC \ref{FPSLE} and the short and long time limits of this equation is derived in this section.

\section{SLE}\label{SLE}
As a well-known fact, the critical 2D models have special algebraic and geometrical properties. The algebraic properties of these models are described within the conformal field theories. However these theories are unable to uncover the geometrical features of these models since it concerns the local fields defined in these models. According to Schramm's idea one can describe the interfaces of two-dimensional (2D) critical models via growth processes named as SLE. Thanks to this theory however, a deep connection between their local properties and their global (geometrical) features has been discovered. This theory aims to describe the 2D critical models by focusing their interfaces. These non-intersecting \textit{domain-walls} are assumed to have two essential properties, conformal invariance and the domain Markov property. In the SLE theory one replaces the critical curve by a dynamical one. \\
We consider the model on the upper half plane, i.e. $H=\left\lbrace z\in \mathbb{C}, \Im z\geq 0\right\rbrace$. Let us denote the curve up to time $t$ as $\gamma_t$ and the hull $K_t$ as the set of points which are located exactly on the $\gamma_t$ trace, or are disconnected from the infinity by $\gamma_t$. The complement of $K_t$ is $H_{t}:=H\backslash{K_{t}}$ which is simply-connected. According to Riemann mapping \cite{Riemann} theorem there is always a conformal mapping $g_t(z)$ (in two dimensions) which maps $H_{t}\rightarrow{H}$. The map $g_{t}(z)$ (commonly named as uniformizing map, meaning that it does uniformize the $\gamma_t$ trace to the real axis) is the unique conformal mapping with $g_{t}(z)=z+\frac{2t}{z}+O(\frac{1}{z^{2}})$ as $z\rightarrow{\infty}$ known as hydrodynamical normalization. Loewner showed that this mapping satisfies the following equation \cite{cardy,Loewner,Smirnov}:
\begin{equation}
\partial_{t}g_{t}(z)=\frac{2}{g_{t}(z)-\zeta_{t}},
\label{Loewner}
\end{equation}
with the initial condition $g_{t}(z)=z$ and for which the tip of the curve (up to time $t$) is mapped to the point $\zeta_t$ on the real axis. For fixed $z$, $g_{t}(z)$ is well-defined up to time $\tau_{z}$ for which $g_{t}(z)=\zeta_{t}$. The more formal definition of hull is therefore $K_{t}=\overline{\lbrace z\in H:\tau_{z}\leq t \rbrace}$. For more information see \cite{cardy,Loewner,Smirnov,RohdeSchramm}. For the critical models, it has been shown \cite{Shramm} that $\zeta_{t}$ (referred to as the driving function) is a real-valued function proportional to the one-dimensional Brownian motion $\zeta_{t}=\sqrt{\kappa}B_{t}$ in which $\kappa$ is known as the diffusivity parameter. SLE aims to analyze these critical curves by classifying them to the one-parameter classes represented by $\kappa$.\\
There are three phases for these curves:\\
\textbf{Dilute phase}: for $2\leq\kappa\leq{4}$ the trace is non-self-intersecting and it does not hit the real axis. Therefore in this case the hull and the trace are identical: $K_{t}=\gamma_{t}$.\\
\textbf{Dense phase}: for $4<\kappa\leq{8}$, the trace touches itself and the real axis so that a typical point is surely swallowed as $t\rightarrow\infty$ and $K_{t}\neq\gamma_{t}$.\\
\textbf{Space-filling phase}: for $\kappa>8$ the curve is space-filling.\\
There is a connection between the two phases: in the dense phase the frontier of $K_{t}$ (the boundary of $H_{t}$ minus any portions of the real axis) is a SLE curve. It has been shown \cite{cardy} that this simple curve is locally described by the diffusivity parameter $\tilde{\kappa}=\frac{16}{\kappa}$ in the dilute phase.\\
The correspondence to CFT is via a simple relation between diffusivity parameter $\kappa$ and the CFT-central charge $c$, i.e. $c=\frac{(3\kappa-8)(6-\kappa)}{2\kappa}$ \cite{BauBer}.
\subsection{SLE($\kappa,\rho_c$)}\label{SLE(k,r)}
\begin{figure}
\centerline{\includegraphics[scale=.40]{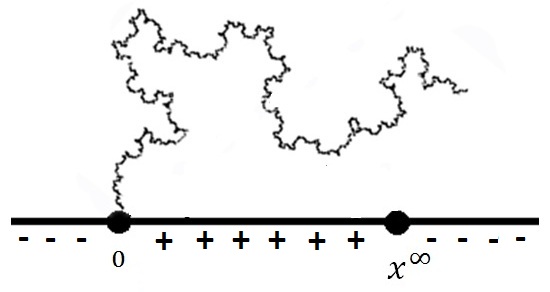}}
\caption{Schematic plot of the SLE trace in presence of a bcc point $x^{\infty}$ (boundary change has been shown Schematically by ($+$) and ($-$)).}
\label{SLEkrEvol1}
\end{figure}
\begin{figure}
\centerline{\includegraphics[scale=.40]{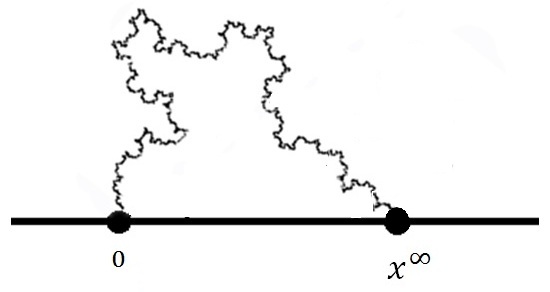}}
\caption{Schematic plot of the SLE trace which ends on the real axis at $x^{\infty}$.}
\label{SLEkrEvol2}
\end{figure}
Let us now suppose that there is a preferred point on the real axis, namely at $x^{\infty}$ as indicated in FIG \ref{SLEkrEvol1} and \ref{SLEkrEvol2}. We are going to analyze the random curve in presence of this point. Note that the conformal invariance of the curve has been explicitly broken, however the curve is yet self-similar. SLE($\kappa,\rho_c$) is a variant of ordinary (chordal) SLE which analyzes such curves. We consider the upper half plane as above. In this theory the parameter $\kappa$ is the same as above. The parameter $\rho_c$ has to do with the boundary conditions (bc) imposed. The point $x_{\infty}$ on the real axis mark the location where boundary condition changes (we name it as $x_{\infty}$ since this point may be considered as the point at which the random curve returns as $t\rightarrow \infty$, see below).  The exact value of $\rho_c$ depends on the boundary change at this preferred point. In the corresponding CFT, the bcc imposed in $x_{\infty}$ corresponds to putting a bcc operator with conformal weight $h_{\rho_c}=\frac{\rho_c \left( \rho_c+4-\kappa\right)}{4\kappa}$ at that point \footnote{Another bcc operator should be put at the origin with the conformal weight $h_o=\frac{6-\kappa}{2\kappa}$} \cite{NajafiLPP,Kytola}.\\
For these curves the stochastic equation is the same as formula (\ref{Loewner}). The driving function $\zeta_t$ acquires however a drift term with respect to the chordal SLE: 
\begin{equation}
d\zeta_{t}=\sqrt{\kappa}dB_{t}+\frac{\rho_c}{\zeta_{t}-x_t^{\infty}}dt
\label{SLE(k,r)driving}.
\end{equation}
in which $B_t$ is one-dimensional Brownian motion and $x_t^{\infty} \equiv g_{t}(x_{\infty})$ is the transformation of the point $x_{\infty}$ under the action of $g_t$. This situation can be generalized to a set up with the arbitrary number of bcc's.\\
\\
\textbf{Real-axis to real-axis random curves:}\\
As an example of the above discussion, we consider the FIG \ref{SLEkrEvol2} in which the mentioned preferred point is the ending point of the curve (which does not go to the infinity, but after the exploration process turns to the real axis). For this set-up, by using the conformal map $\phi={x_{\infty}z}/{(x_{\infty}-z)}$, one can send the end point of the curve to the infinity. Since our model is conformal invariant, we can deduce that the function $h_{t}=\phi{\circ }g_{t}{\circ}\phi^{-1}$ describes ordinary (chordal) SLE from which one easily concludes that $g_{t}$ satisfies $\partial_{t}g_{t}=2/(\lbrace{\phi^{\prime}(g_{t})(\phi(g_{t})-\zeta_{t})}\rbrace)$ in which $\phi^{\prime}(z)\equiv \frac{\text{d}\phi}{\text{d}z}$. The other criteria is to make the mapping hydrodynamically normalized. It has been shown \cite{BauBer2} that the stochastic equation of hydrodunamically normalized $g_{t}$ is the same as Eq. (\ref{Loewner}) with the price that $\zeta_t$ acquires a drift term as follows \cite{BauBer2}:
\begin{equation}
d\zeta_{t}=\sqrt{\kappa}dB_{t}+\frac{\kappa-6}{\zeta_{t}-g_{t}(x_{\infty})}dt
\label{driving}
\end{equation}
In the other words, this stochastic function is the driving function of the SLE($\kappa,\rho_c$) with $\rho_c=\kappa-6$. Thus for the critical curves from boundary to boundary, the corresponding driving function acquires a drift term. This generalization of SLE can be generalized further to have multiple preferred real axis points and multi-critical curves \cite{BerBau}. For review see also references \cite{cardy,RohdeSchramm}. \\
The other example is the traces which terminate surely to the real axis in the interval $[x_0,\infty)$. For these curves of course the point $x_0$ is regarded as the preferred point for which $\rho_c=(\kappa-6)/2$ yields dipolar SLE($\kappa$) \cite{BauBer,BerBau,BauBer2}.
\section{Derivation of Fokker-Planck Equation}\label{FP}
\begin{figure}
\centerline{\includegraphics[scale=.40]{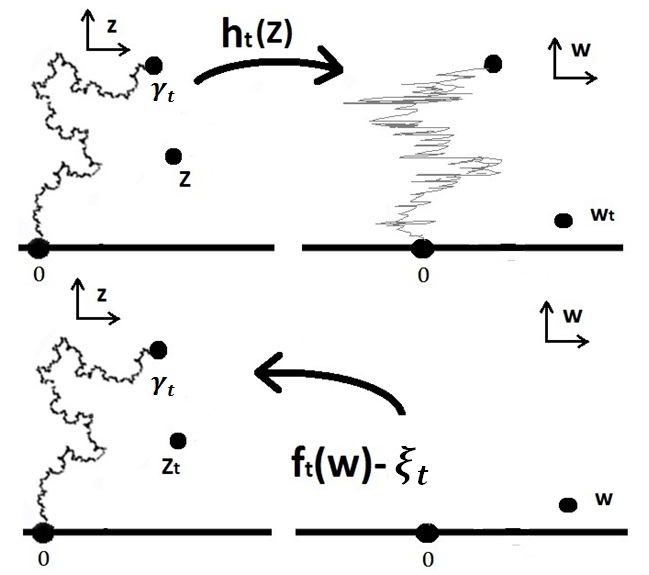}}
\caption{Schematic representation of the conformal map $h_t(z)$ and the backward conformal map $f_t(w)-\zeta_t$ on the SLE trace.}
\label{mappings}
\end{figure}
\begin{figure}
\centerline{\includegraphics[scale=.40]{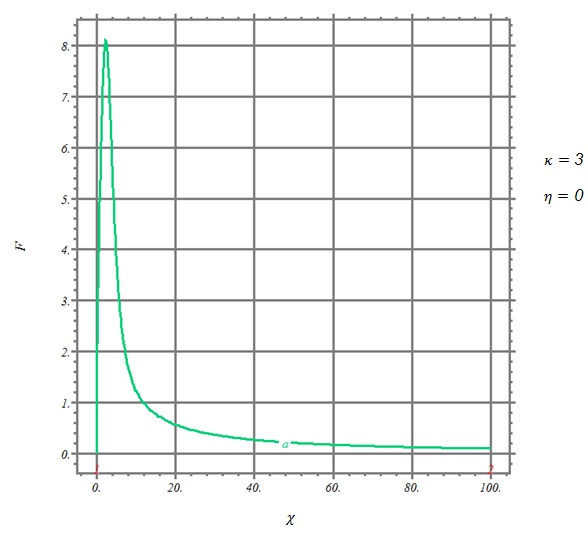}}
\caption{The solution of the equation \ref{FPE-t-indep} for the SLE curve with diffusivity parameter $\kappa=3$. In these graphs $F$ has been plotted in terms of $\chi$ ($\eta=0$).}
\label{FPESolution1}
\end{figure}
\begin{figure}
\centerline{\includegraphics[scale=.40]{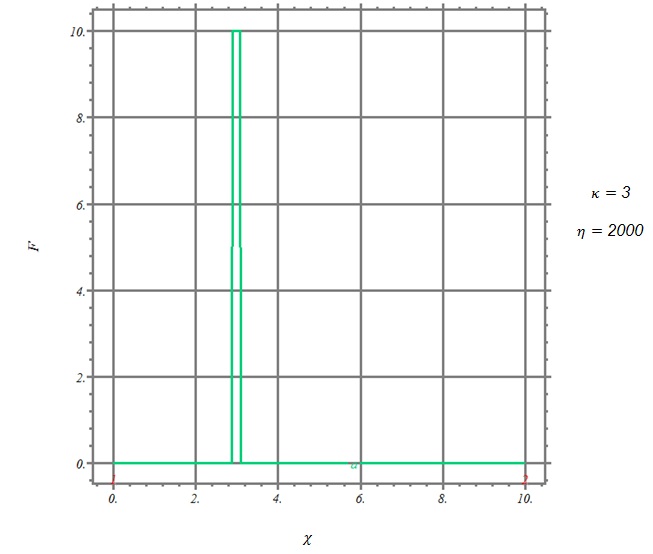}}
\caption{The solution of the equation \ref{FPE-t-indep} for the SLE curve with diffusivity parameter $\kappa=3$. In these graphs $F$ has been plotted in terms of $\chi$ ($\eta=1000$).}
\label{FPESolution2}
\end{figure}
\begin{figure}
\centerline{\includegraphics[scale=.40]{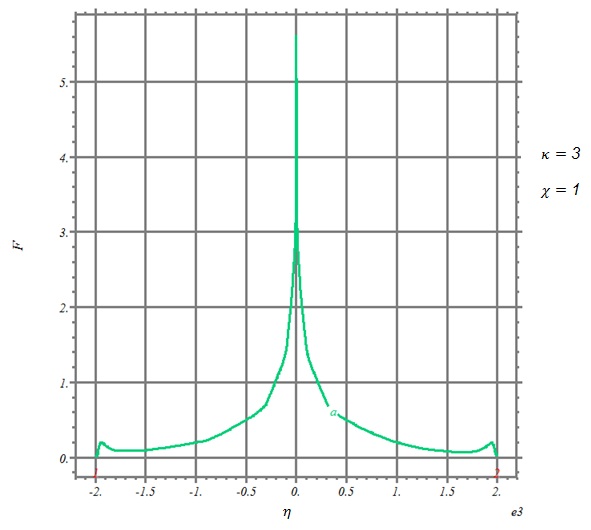}}
\caption{The solution of the equation \ref{FPE-t-indep} for the SLE curve with diffusivity parameter $\kappa=3$. In these graphs $F$ has been plotted in terms of $\eta$ ($\chi=1$).}
\label{FPESolution3}
\end{figure}
\begin{figure}
\centerline{\includegraphics[scale=.40]{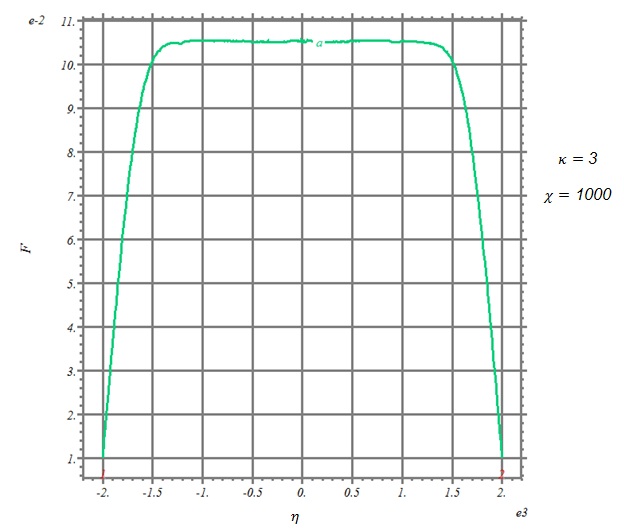}}
\caption{The solution of the equation \ref{FPE-t-indep} for the SLE curve with diffusivity parameter $\kappa=3$. In these graphs $F$ has been plotted in terms of $\eta$ ($\chi=100$).}
\label{FPESolution4}
\end{figure}

In this section we are going to find the Langevin equation for the tip of the chordal SLE trace. Let us first define the shifted conformal map as follows:
\begin{equation}
h_t(z)\equiv g_t(z)-\zeta_t
\end{equation}
for which one can easily verify that
\begin{equation}
h_t^{-1}(w)\stackrel{d}{=}g_t^{-1}(w+\zeta_t).
\label{equivlence}
\end{equation}
In the above equation $\stackrel{d}{=}$ means the equality of the distributions of stochastic processes. The differential equation governing $h_t$ is:
\begin{equation}
\begin{split}
\partial_th_t(z)=\frac{2}{h_t(w)}-\partial_t\zeta_t
\end{split}
\label{h_eq}
\end{equation}
in which $h_0(z)=z$. One can retrieve the SLE trace by the relation $\gamma(t)=\lim_{\epsilon\downarrow{0}}g_{t}^{-1}(\zeta_{t}+i\epsilon)=\lim_{z\rightarrow \zeta_t}g_t^{-1}(z)$ and find the differential equation of the tip of SLE trace. The equation governing $g_t^{-1}$ is very difficult to solve. There is a way out of this problem, using the backward SLE equation. It is defined as follows (note that $\zeta_t\stackrel{d}{=}\zeta_{-t}\stackrel{d}{=}-\zeta_t$) \cite{Gruzberg}:
\begin{equation}
\begin{split}
\partial_tf_t(w)=-\frac{2}{f_t(w)-\zeta_t}
\end{split}
\label{backw}
\end{equation}
It has been shown that the probability distribution of $f_t$ is the same as $g_t^{-1}$ \cite{Gruzberg1}, i.e.
\begin{equation}
\begin{split}
f_t(w)-\zeta_t\stackrel{d}{=} g_t^{-1}(w+\zeta_t)\stackrel{d}{=}h_t^{-1}(w).
\end{split}
\label{equivlence2}
\end{equation}
The schematic representation of this equation has been sketched in FIG \ref{mappings}. Therefore the tip of the SLE trance can be obtained $\gamma_T=\lim_{w\rightarrow 0}g_T^{-1}(w+\zeta_T)=\lim_{w\rightarrow 0}h_T^{-1}(w)$. Now we can find the trajectory of the tip of the SLE trace using the backward equation (\ref{backw}) for $z_t=x_t+y_t$ which is (notice that Re$(\gamma_t)$ and Im$(\gamma_t)$ have the same joint distribution as $x_t$ and $y_t$)
\begin{equation}
\begin{split}
& \partial_tx_t=-\frac{2x_t}{x_t^2+y_t^2}-\partial_t\zeta_t\\
& \partial_ty_t=\frac{2y_t}{x_t^2+y_t^2}
\end{split}
\label{BKWComponents}
\end{equation}
conditioned to have the initial values $x_0=u$ and $y_0=v$ in which $w=u+iv$ is the initial point of the flow. 	
The Fokker-Planck equation for SLE governs the probability density $\rho(x,y,t)$ for the critical curves which is the probability that the tip of the curve be at the point $(x,y)$ on the upper half plane at the time $t$. It is defined as $\rho(x,y,t)=\langle \delta(x-x_t)\delta(y-y_t)\rangle$. To obtain the FP equation, we notice that $\rho(x+dx,y+dy,t+dt)=\rho(x,y,t)$. Using the Langevin equations (\ref{BKWComponents}) of the growing SLE curve and after performing some Ito calculations, noticing that for the chordal SLE, the driving function is proportional to the Brownian motion (which implies that $\left\langle (d\zeta_t)^2\right\rangle = \kappa dt$), it can be easily proved that \cite{Gruzberg}:
\begin{equation}
\partial_t \rho(x,y,t)=\left[\frac{\kappa}{2}\partial_x^2 +\frac{2(x-\zeta_0)}{(x-\zeta_0)^2+y^2}\partial_x -\frac{2y}{(x-\zeta_0)^2+y^2}\partial_y \right]\rho(x,y,t)
\label{FPE}
\end{equation}
in which $\zeta_0$ is the point at which the curve starts to grow. This equation has the symmetry $(x-\zeta_0) \rightarrow \lambda (x-\zeta_0)$ , $y\rightarrow \lambda y$ and $t\rightarrow \lambda^2 t$ in which $\lambda$ is some (non-zero) positive real parameter. One can see this symmetry in the other way. Let us observe the effect of the transformation $x \rightarrow \lambda x$ , $y\rightarrow \lambda y$ (setting $\zeta_0=0$) on $\rho(x,y,t)$ which yields $\left\langle \delta(\lambda x-x_t)\delta(\lambda y-y_t)\right\rangle=\frac{1}{\lambda^2}\left\langle \delta(x-\frac{1}{\lambda}x_{\lambda^2 T})\delta( y-\frac{1}{\lambda}y_{\lambda^2 T})\right\rangle$ in which $T=\lambda^{-2} t$. Noting that $\frac{1}{\lambda}x_{\lambda^2T}\stackrel{d}{=}x_T$ and $\frac{1}{\lambda}y_{\lambda^2T}\stackrel{d}{=}y_T$ (see Eq. \ref{BKWComponents}), we find that
\begin{equation}
\rho(\lambda x,\lambda y, \lambda^2 t)=\frac{1}{\lambda^2}\rho(x,y,t).
\end{equation}
Considering also the translational symmetry of the problem, i.e. $x\rightarrow x+a$ and $\zeta_0\rightarrow \zeta_0+a$, we expect its general form to be:
\begin{equation}
\rho(x,y,t)=r^{-2}g\left(\chi_t,\eta\right)=t^{-1}f\left(\chi_t,\eta\right)
\label{scaling}
\end{equation}
in which $\chi_t\equiv\frac{(x-\zeta_0)^2+y^2}{t}$, $\eta\equiv \frac{x-\zeta_0}{y}$, $\alpha$,  $r\equiv \sqrt{(x-\zeta_0)^2+y^2}$ and $f$ and $g$ are some functions to be determined by Eq [\ref{FPE}] (not to be confused with mapping $f_t$ defined before). By setting Eq \ref{scaling} into Eq \ref{FPE} we see that it is reduced to the following form:

\begin{equation}
\begin{split}
& h_{\chi\chi}\partial_{\chi}^2f+2h_{\chi\eta}\partial_{\chi}\partial_{\eta}f+h_{\eta\eta}\partial_{\eta}^2f+h_{\chi}\partial_{\chi}f+h_{\eta}\partial_{\eta}f+f=0\\
& h_{\chi\chi}=2\kappa\frac{\eta^2\chi}{1+\eta^2}\\
& h_{\eta\eta}=\frac{\kappa}{2}\frac{(1+\eta^2)}{\chi}\\
& h_{\chi\eta}=\kappa\eta\\
& h_{\chi}=\kappa+\chi+\frac{4(\eta^2-1)}{\eta^2+1}\\
& h_{\eta}=4\frac{\eta}{\chi}
\end{split}
\label{FPE-t-indep}
\end{equation}
It can easily be checked that this equation is a hyperbolic PDE, i.e. $h_{\chi\eta}^2-h_{\chi\chi}h_{\eta\eta}=0$. By defining $\xi\equiv \frac{1+\eta^2}{\chi}=\frac{t}{y^2}$ (so that $x-\xi_0=\eta\sqrt{\frac{t}{\xi}}$ and $y=\sqrt{\frac{t}{\xi}}$), and $f(\chi,\eta)=F(\xi,\eta)$ we obtain the canonical form:
\begin{equation}
\frac{1}{2}\kappa\xi\partial_{\eta}^2F-\left(\kappa+\frac{1+\eta^2}{\xi}+4\frac{\eta^2-1}{\eta^2+1}\right)\left(\frac{\xi^2}{1+\eta^2}\right)\partial_{\xi}F+\frac{4\eta\xi}{1+\eta^2}\partial_{\eta}F+F=0
\label{FPE2}
\end{equation}
Before investigating this equation, let us explore the properties of the regime of our interest, i.e. long (non-primitive) times. In this limit we expect that $x_t\gg y_t$. Consequently we are interested in the large $\eta$ limit of the above equation. In this limit we have:

\begin{equation}
\frac{1}{2}\kappa\partial_{\eta}^2F-\partial_{\xi}F+\frac{4}{\eta}\partial_{\eta}F+\frac{F}{\xi}=0
\label{FPE-longTimes}
\end{equation}
To solve this equation we consider the factored form of solution, i.e. $F=\Pi(\eta)\Sigma(\xi)$ from which we conclude that:

\begin{equation}
\left\lbrace \begin{array}{c} \partial_{\xi}\Sigma-(\lambda+\frac{1}{\xi})\Sigma=0 \\  \frac{1}{2}\kappa\partial_{\eta}^2\Pi+\frac{4}{\eta}\partial_{\eta}\Pi-\lambda\Pi=0 \end{array}\right.
\end{equation}
In this equation $\lambda$ is an arbitrary real number. The solution of the above equations are 

\begin{equation}
\begin{split}
& \Sigma(\xi)=\Sigma_0\left(\frac{\xi}{\xi_0}\right)e^{\lambda(\xi-\xi_0)}\\  &\Pi(\eta)=\eta^{-\frac{8-\kappa}{2\kappa}}I_{\frac{\kappa-8}{2\kappa}}\left(i\sqrt{\frac{2\lambda}{\kappa}}\eta\right)+\eta^{-\frac{8-\kappa}{2\kappa}}K_{\frac{\kappa-8}{2\kappa}}\left(i\sqrt{\frac{2\lambda}{\kappa}}\eta\right)
\end{split}
\end{equation}
in which $I$ and $K$ are modified Bessel functions of first and second kind respectively. The expected behavior of these functions (being real in each interval) is satisfied only when $\lambda=0^-$ which yields $\Pi(\eta)=\eta^{-\frac{8-\kappa}{\kappa}}$ and causes $\Sigma(\xi)$ to have a maximum at some point and uniformly go to zero at large values of $\xi$. Therefore for long times we obtain that
\begin{equation}
\rho(x,y,t)\sim\frac{\rho_0}{t}\frac{\Sigma(\xi)}{\eta^{(8-\kappa)/(\kappa)}}=\frac{\rho_0}{t}\left(\frac{y}{x}\right)^{(8-\kappa)/(\kappa)}\Sigma(\xi)
\label{longTimeRho}
\end{equation}
in which $\rho_0$ is the proportionality constant. This result reveals some important behaviors. For $\kappa<8$ the probability of touching the real axis boundary, i.e. $\eta=\frac{x}{y}\rightarrow \pm\infty$ goes to zero as $\frac{1}{\eta^{(8-\kappa)/\kappa}}$, but for $\kappa=8$ there is a finite (maximum) probability of touching the boundary which is an important effect in SLE, having its root in the fact that for this case the curve is space-filling. For large amounts of $\eta$ also, $\rho(x,y,t)\sim \lim_{\lambda\rightarrow 0^-}\xi e^{\lambda\xi}$ goes to zero for large $\xi$ values.\\
To be self-contained, we have numerically solved the equation \ref{FPE-t-indep}. To this end we have used the finite difference method. We have considered a lattice with dynamic mesh sizes (bounded to $0\leq\chi\leq 10^3$ and $-2\times10^3\leq\eta\leq 2\times10^3$). Due to the singular behavior of Eq. (\ref{FPE-t-indep}) near $\chi=0$ and $\eta=0$, one should mesh the space near these lines more exact (smaller) than the other points \footnote{The MATLAB PDE toolbox has been used to solve this equation.}. The results have been indicated in Figs. \ref{FPESolution1},\ref{FPESolution2},\ref{FPESolution3},\ref{FPESolution4}. As is seen in Fig. \ref{FPESolution1}, when $\eta=0$ ($x\rightarrow 0$ and $t=$ finite), $F$ has a single peak at some $\chi$. In fact we have observed that these peaks occur at $\chi_0=\frac{x_0^2+y_0^2}{t}\simeq\frac{y_0^2}{t}=a(\kappa)$ in which $a(\kappa)\sim (\kappa)^{-\frac{1}{2}}$. This fact shows that for larger amounts of $\kappa$, at a fixed time, the vertical growth (along the $y$ axis) of the growing curve decreases. Figure \ref{FPESolution2} shows this dependence for large $\eta$, i.e. $\left|x\right|\gg y$ and $t=$ finite. The figure shows a very narrow peak appears around $\chi_0\simeq \frac{x_0^2}{t}=b(\kappa)$ in which $b(\kappa)\sim \kappa$ which is expected, since for long times $x_t\simeq \sqrt{\kappa}B_t$ (Eq. \ref{BKWComponents}). The symmetry of $F$ with respect to $\eta$ is evident in Figs. \ref{FPESolution3} and \ref{FPESolution4} in which the behavior of $F$ has been sketched in terms of $\eta$ for the cases $\chi=1$ and very large amount of $\chi$ respectively. For $\chi=1$ ($r^2=t$) it is shown that $F$ has a peak around $\eta=0$ and decreases (for positive $\eta$ axis) as $\eta$ increases and for $\eta\rightarrow \infty$ (2000 in the graph) it tends to zero according to the relation obtained above. For $\chi\gg 1$ (1000 in the Fig. \ref{FPESolution4}) the sensitivity of $F$ to $\eta$ ($=\cot\theta$) becomes negligible and its absolute value becomes very small.\\
Before closing this sub-section, we notice that from the general solution of $F(\xi,\eta)$, we can also calculate $\left\langle x^n\right\rangle $ and $\left\langle y^n\right\rangle$ as follows:

\begin{equation}
\begin{split}
\left\langle x^n\right\rangle(t)&=\int_{-\infty}^{\infty}\text{d}x\int_{0}^{\infty}\text{d}y x^n\rho(x,y,t)\\
&=\frac{1}{2}t^{\frac{n}{2}}\int_{-\infty}^{\infty}\text{d}\eta\int_{0}^{\infty}\text{d}\xi \frac{ \eta^nF(\xi,\eta)} {\xi^{\frac{n}{2}+2}}
\end{split}
\label{xy_t1}
\end{equation}
\begin{equation}
\begin{split}
\left\langle y^n\right\rangle(t)&=\int_{-\infty}^{\infty}\text{d}x\int_{0}^{\infty}\text{d}y y^n\rho(x,y,t)\\
&=\frac{1}{2}t^{\frac{n}{2}}\int_{-\infty}^{\infty}\text{d}\eta\int_{0}^{\infty}\text{d}\xi \frac{F(\xi,\eta)} {\xi^{\frac{n}{2}+2}}
\end{split}
\label{xy_t2}
\end{equation}
\begin{equation}
\begin{split}
\left\langle x^ny^m\right\rangle(t)&=\int_{-\infty}^{\infty}\text{d}x\int_{0}^{\infty}\text{d}y x^ny^m\rho(x,y,t)\\
&=\frac{1}{2}t^{\frac{1}{2}(n+m)}\int_{-\infty}^{\infty}\text{d}\eta\int_{0}^{\infty}\text{d}\xi \frac{\eta^nF(\xi,\eta)} {\xi^{\frac{1}{2}(n+m)+2}}
\end{split}
\label{xy_t3}
\end{equation}
We see that the $n$-th moment of displacement scales as $\left\langle r^n\right\rangle \sim t^\frac{n}{2}$. We can learn more about $F$ from these relations. Since Eq \ref{FPE2} is invariant under $\eta\rightarrow -\eta$, one can conclude that its solutions are classified into even and odd solutions with respect to the $\eta$ component. By setting $n=1$ in Eq \ref{xy_t1}, and using the fact that $\left\langle x\right\rangle=0$ we can easily find that only even solutions should be taken into account, i.e. $F(\xi,\eta)$ is even in $\eta$ component (note that we can not directly generalize this conclusion to the limiting case $F\simeq \eta^{-(8-\kappa)/(\kappa)}$. In fact as we see later this function should be of the form $(1+\eta^2)^{-(8-\kappa)/(2\kappa)}$ which is even function of $\eta$ and for large amounts of $\eta$ becomes the above result).\\
These results are valid within the Fokker-Planck equation and should be verified by the Langevin equations \ref{BKWComponents} which is the subject of the next sub-section.
\subsection{Langeving equation for the long-time limit}
In this sub-section we obtain the behavior of $\bar{x}(t)\equiv\sqrt{\left\langle x^2\right\rangle(t)}$ and $\left\langle y\right\rangle(t)$ for long times, using the Langevin equation ( Eq. [\ref{BKWComponents}]). In this limit we have $\frac{x_t}{x_t^2+y_t^2}\rightarrow 0$ and therefore $x_t\sim \sqrt{\kappa} B_t$. Therefore $\bar{x}(t)\simeq \sqrt{\kappa}t^{\frac{1}{2}}$ in accordance with Eq. \ref{xy_t1}. To find the behavior of $y_t$, we notice that in long-time limit, the probability distribution for $x_t\stackrel{d}{=}\sqrt{\kappa}B_t$ is $P(x,t)=\frac{1}{\sqrt{2\pi\kappa t}}e^{-\frac{x^2}{2\kappa t}}$. Using the second line of Eq \ref{BKWComponents} and replacing the r.h.s with its average, we obtain:
\begin{equation}
\begin{split}
& \partial_ty_t\simeq \frac{1}{\sqrt{2\pi\kappa t}}\int_{-\infty}^{\infty}\frac{2y_t}{y_t^2+x^2}e^{-\frac{x^2}{2\kappa t}}\text{d}x\\
\Rightarrow & e^{-\chi_t^2}\left((\kappa t)^{\frac{1}{2}}\partial_t(\sqrt{\kappa t}\chi_t)\right)\simeq 2\int_{0}^{\chi_t}e^{-\eta^2}\text{d}\eta
\end{split}
\label{y_tLarge}
\end{equation}
in which $\chi_t\equiv \frac{y_t}{\sqrt{2\kappa t}}$. One can easily obtain the differential equation governing $\chi_t$, i.e.
\begin{equation}
\begin{split}
\partial_t\left( t\partial_t\chi_t\right)-2t\chi_t(\partial_t\chi_t)^2-\left(\frac{4-\kappa}{2\kappa}+\chi_t^2\right)\partial_t\chi_t=0
\end{split}
\label{y_tLarge2}
\end{equation}
If we represent this function again as $\chi_t=At^{-\frac{1}{2}}f(t)$ (in which $f(t)$ is proportional to $y_t$), then the following equation is obtained ($f^{\prime}\equiv \frac{\text{d}}{\text{d}t}f(t)$ and $f^{\prime\prime}\equiv \frac{\text{d}^2}{\text{d}t^2}f(t)$):
\begin{equation}
\begin{split}
\frac{1}{4}At^{-\frac{3}{2}}f+At^{\frac{1}{2}}f^{\prime\prime}-2A^3t^{-\frac{5}{2}}f\left( -\frac{1}{2}f+tf^{\prime}\right)^2-At^{-\frac{3}{2}}\left(\frac{4-\kappa}{2\kappa}+A^2t^{-1}f^2\right)\left( -\frac{1}{2}f+tf^{\prime}\right)=0
\end{split}
\label{y_tLarge3}
\end{equation}
It can easily be checked that the answer of this equation is $f(t)\sim t^{\frac{1}{2}}$. The conclusion is that in the long-time limit we have:
\begin{equation}
\left\lbrace \begin{array}{c} x_t\stackrel{d}{=} \sqrt{\kappa}B_t \sim t^{\frac{1}{2}} \\  y_t\sim t^{\frac{1}{2}}\end{array}\right.
\end{equation}
which is in agreement with the results of the previous sub-section. 
\subsection{CFT background}\label{CFTchordal}
We have found the solution of the FPE of SLE growing curves in long-time limit in the previous sections. There is another way for exploring this problem within conformal filed theory (CFT) techniques which is the subject of this section.\\
Let us define the function $P(x,y)$ as the probability distribution of curves passing the region $(x,x+dx)$ and $(y,y+dy)$. It is defined as:
\begin{equation}
P(x,y)=\int_0^{\infty}\rho(x,y,t)\text{d}t=\int_0^{\infty}\left\langle \delta(x-x_t)\delta(y-y_t)\right\rangle \text{d}t
\label{probdens1}
\end{equation}
Throughout of this sub-section we name the point at which critical curve starts to grow as $\zeta_0$ which is important in our analysis. The behavior of this function under the transformation $(x-\zeta_0) \rightarrow \lambda (x-\zeta_0)$ , $y\rightarrow \lambda y$ is obvious:

\begin{equation}
\begin{split}
P(x,y)|_{(x-\zeta) \rightarrow \lambda (x-\zeta), y\rightarrow \lambda y}&\longrightarrow\int_0^{\infty}\rho(\lambda x,\lambda y,t)\text{d}t\\
&=\frac{1}{\lambda^2}\int_0^{\infty}\rho(x,y,T)\times\lambda^2\text{d}T=P(x,y)
\end{split}
\label{probdens2}
\end{equation}
and its more explicit form is:
\begin{equation}
\begin{split}
P(x,y)&=\int_0^{\infty}t^{-1}F\left(\xi_t,\eta\right)\text{d}t\\
&=\int_0^{\infty}\frac{F(\xi,\eta)}{\xi}\text{d}\xi
\end{split}
\label{probdens3}
\end{equation}
showing that it has zero dimension. \\
Let us investigate the properties of $P(x,y,\epsilon)$ which is the probability that the distance of the point $z=x+iy$ from the SLE curve be $\epsilon$. In the other words, it is the probability that the curve passes a disc with radius $\epsilon$ centered at $z=x+iy$. This probability is compatible with $P(x,y)$ with a finite area (disc with radius $\epsilon$) around. From the CFT point of view and noting that the original model has the conformal symmetry, one can easily understand that $P(x,y,\epsilon)$ is proportional to the following multi-point correlation function:
\begin{equation}
P(z,\bar{z},\zeta_0,x^\infty)=\frac{\left\langle \hat{F}(x,y)\hat{F}(x,-y)\psi(\zeta_0)\psi(x^\infty)\right\rangle}{\left\langle \psi(\zeta_0)\psi(x^\infty)\right\rangle}.
\label{corr}
\end{equation}
In this equation, $\hat{F}(x,y)$ is a \textit{detector operator} whose value is non-zero in a field configuration only if the mentioned curve (defined in the background of other fields) passes the infinitesimal region around the point $(x,y)$ and $x^{\infty}$ is the point on the boundary at which the SLE curve ends (in other words, we have two boundary changes, one at $\zeta_0$ and the other at $x^{\infty}$). The operator $\hat{F}(x,-y)$ is the image of $\hat{F}(x,y)$ with respect to the real axis whose presence in the correlators in the boundary CFTs (BCFTs) is customary. $\psi$ is the boundary changing operator for the CFT corresponding to underling SLE whose conformal weight is $h_1(\kappa)=\frac{6-\kappa}{2\kappa}$ with second level null vector \cite{BerBau}:
\begin{equation}
\left( \frac{\kappa}{2}L_{-1}^2-2L_{-2}\right) \psi=0
\label{virasoro}
\end{equation}
in which $L_{n}$'s are the generators of the Virasoro algebra satisfying
\begin{equation}
\left[L_n,L_m\right]=(n-m)L_{n+m}+\frac{c}{12}n(n^2-1)\delta_{n+m,0}.
\end{equation}
and $c$ is the central charge of the underlying CFT. Using Ward identities, Eq. \ref{virasoro} leads to the following equation for the mentioned correlation function \cite{Dif}:
\begin{equation}
\left( \frac{\kappa}{2}\mathcal{L}_{-1}^2-2\mathcal{L}_{-2}\right)\left\langle \hat{F}(x,y)\hat{F}(x,-y)\psi(\zeta_0)\psi(x^\infty)\right\rangle=0
\label{f_dif}
\end{equation}
with
\begin{equation}
\begin{split}
&\mathcal{L}_{-1}=\frac{\partial }{\partial \zeta_0}\\
\mathcal{L}_{-2}&=\sum_i \left( \frac{h_i}{(z_i-\zeta_0)^2}-\frac{1}{z_i-\zeta_0} \frac{\partial}{\partial z_i} \right).
\end{split}
\end{equation}
in which the sum is over each field in the correlation function Eq. (\ref{corr}) except $\zeta_0$. From global conformal invariance we know that
\begin{equation}
Q(z,\bar{z},\zeta_0,x^\infty)\equiv\left\langle \hat{F}(x,y)\hat{F}(x,-y)\psi(\zeta_0)\psi(x^\infty)\right\rangle=\frac{1}{\left( \zeta_0-x^\infty\right)^{2h_1(\kappa)}}P(z,\bar{z},\zeta_0,x^\infty)
\label{Q-P}
\end{equation} 
so that $Q$ satisfies the equation
\begin{equation}
\left[ \frac{\kappa}{4}\partial_{\zeta_0}^2- 2h_F\textrm{Re}\left( \frac{1}{(z-\zeta_0)^2} \right)+\frac{1}{z-\zeta_0} \frac{\partial}{\partial z}+\frac{1}{\bar{z}-\zeta_0} \frac{\partial}{\partial \bar{z}}+\frac{1}{x^\infty-\zeta_0} \frac{\partial}{\partial x^\infty}- \frac{h_1(\kappa)}{(x^\infty-\zeta_0)^2}\right] Q=0.
\label{LPE}
\end{equation}
In this equation $h_F$ is conformal weight of $\hat{F}$. From general grounds of CFT, we can easily find that the four-point correlation function in Eq \ref{corr} has the following form:
\begin{equation}
\begin{split}
P&=y^{-\frac{4}{3}h_F+\frac{2}{3}h_1(\kappa)}(x^\infty-\zeta_0)^{\frac{2}{3}h_1(\kappa)+\frac{2}{3}h_F}[(x-x^\infty)^2+y^2]^{\frac{-1}{3}h_1(\kappa)-\frac{1}{3}h_F}[(x-\zeta_0)^2+y^2]^{\frac{-1}{3}h_1(\kappa)-\frac{1}{3}h_F}h_0(u)\\
&=\frac{1}{y^{2h_F}}h(u)
\end{split}
\label{FPF}
\end{equation}
in which $u\equiv \text{Re}\left[\frac{(z-\zeta_0)(\bar{z}-x^{\infty})}{y(x^{\infty}-\zeta_0)}\right]$ ($\eta=\lim_{x^{\infty}\rightarrow \infty}u$) and $h_0$ and $h$ are some functions to be determined. For the limit $x^\infty\rightarrow \infty$ the above equation reduces to the following form:
\begin{equation}
\left[ \frac{\kappa}{4}\partial_{\zeta_0}^2- 2h_F\textrm{Re}\left( \frac{1}{(z-\zeta_0)^2} \right)+\frac{1}{z-\zeta_0} \frac{\partial}{\partial z}+\frac{1}{\bar{z}-\zeta_0} \frac{\partial}{\partial \bar{z}}\right] Q=0.
\label{LPE2}
\end{equation}
and $\text{Re}[u]=\eta$ and $P(z,\bar{z},\zeta_0)\sim Q(z,\bar{z},\zeta_0)=\frac{1}{y^{2h_F}}h(\eta)$. In this case, by substituting Eq \ref{FPF} in Eq \ref{LPE2} we obtain ($\zeta_0=0$):

\begin{equation}
\begin{split}
&\left[ \frac{\kappa}{2}\partial_{x}^2+\frac{2x}{x^2+y^2}\partial_x-\frac{2y}{x^2+y^2}\partial_y+\frac{8h_Fy^2}{\left( x^2+y^2\right)^2}\right]h(u)=0\\
&\Rightarrow \left[\frac{\kappa}{2}\left(1+\eta^2\right)^2\partial_{\eta}^2+4\eta\left(1+\eta^2\right)\partial_{\eta}+8h_F\right]h(\eta)=0
\end{split}
\label{eta_Eq}
\end{equation}
Now we note that the conformal weight of $P(x,y,\epsilon)$ should be zero. The only scale which may compensate the factor $y^{-2h_F}$ in front of $P$ is $\epsilon$ so that it should be of the form $P(x,y,\epsilon)\sim \left(\frac{\epsilon}{y}\right)^{2h_F}h(\eta)\equiv \frac{1}{y^{2h_F}}h(\eta,\epsilon)$ so that $2h_Fh(\eta,\epsilon)=\epsilon\partial_{\epsilon}h(\eta,\epsilon)$ which yields:
\begin{equation}
\left[\frac{\kappa}{2}\left(1+\eta^2\right)^2\partial_{\eta}^2+4\eta\left(1+{\eta}^2\right)\partial_{\eta}+4\epsilon\partial_{\epsilon}\right]h(u,\epsilon)=0
\end{equation}
This equation can be obtained directly by employing the SLE equations, noting that under each conformal transformation $g_{\delta t}(z)$, the mentioned disc is transformed as $\epsilon\rightarrow \left|g_{\delta t}^{\prime}(z)\right|\epsilon\sim (1-2\delta t\text{Re}(1/z^2))\epsilon$ \cite{cardy}. If we use the trial solution $h(\eta)=(1+\eta^2)^{\gamma}$ for the above differential equation, we obtain $h_F=\frac{1}{2}-\frac{1}{16}\kappa$ and $\gamma=\frac{\kappa-8}{2\kappa}$. We know that the fractal dimension of the SLE curve is $d_f=1+\kappa/8$ \cite{cardy}, yielding $2h_F=2-d_f$. Therefore we have:
\begin{equation}
P(x,y,\epsilon)\sim \left(\frac{\epsilon}{y}\right)^{2-d_f}\frac{1}{(1+\eta^2)^{\frac{8-\kappa}{2\kappa}}}=\frac{\epsilon^{2-d_f}}{r^{(8-\kappa)/\kappa}}y^{\frac{1}{8}(\kappa-8)^2}
\end{equation}
First observe that $P\sim r^{-\frac{8-\kappa}{8}}\left(\sin\theta\right)^{-(8-\kappa)/8} \left(1+\cot^2\theta\right)^{-(8-\kappa)/2\kappa}$ (in which $\eta=\cot\theta$) , showing that as $\kappa$ (and also the fractal dimension)  increases, $P$ becomes more long-range. The obvious example is $\kappa=0$ whose curves are straight lines and its probability distribution is expected to have the form $P\sim r^{-1}$. The opposite limit is $\kappa=8$ whose curves are space-filling which leads to the fact that the probability distribution is independent of $r$. The other important result from this formula is that in the limit $y\rightarrow 0$ and $x=\text{finite}$, $P=0$ except for the case $\kappa=8$. Note that this solution is compatible with the one obtained in SEC \ref{FP} in which it was shown that for $\eta\gg 1$, $\Pi(\eta)\sim \eta^{-(8-\kappa)/\kappa}$ which is exactly compatible with the dependence of $P(x,y)$ on $\eta$ obtained above, i.e. $(1+\eta^2)^{-(8-\kappa)/2\kappa}\simeq \eta^{-(8-\kappa)/\kappa}$. This is due to the fact that for large limit of $\eta$, $F$ is separable into $\Pi$ and $\Sigma$ and the integral in Eq \ref{probdens1} $\Pi$ can be brought out of the integral, maintaining its dependence.\\
\section{Fokker-Planck Equation of SLE($\kappa,\rho_c$)}\label{FPSLE}
Now we consider the SLE curves which experience a boundary condition change in some point on the real axis ($x^\infty$). Specifically we analyze the curves which goe from origin to a point on the real axis. As mentioned in section \ref{SLE(k,r)}, these curves are describe by the framework of SLE($\kappa,\kappa-6$) in which we use the relation of the driving function Eq [\ref{driving}]. The Cartesian components of the tip of SLE growing curve satisfy the following Langevin relations:
\begin{equation}
\begin{split}
& dx_t=\frac{-2x_t}{x_t^2+y_t^2}dt-d\zeta_t\\
& dy_t=\frac{2y_t}{x_t^2+y_t^2}dt\\
& d\zeta_t=\sqrt{\kappa}dB_t + \frac{\rho_c}{\zeta_t-x_t^{\infty}}dt
\end{split}
\label{BKWComponentsSLEkr}
\end{equation}
Using this equation and some calculations as above, one can calculate the change of the function $\rho(x,y,t)$. After some Ito calculations it can be proved that:
\begin{equation}
\frac{\partial\rho}{\partial t}=\left[\frac{\kappa}{2}\partial_x^2 +\frac{2x}{x^2+y^2}\partial_x -\frac{2y}{x^2+y^2}\partial_y \right]\rho +\rho_c \partial_x \langle \frac{1}{(\zeta_t - x_t^\infty)}\delta(x-x_t)\delta(y-y_t)\rangle
\label{FPSLEkr}
\end{equation}
in which the second term in the r.h.s. is due to the existence of the preferred point on the real axis. 
\subsection{Perturbative FPE of SLE($\kappa,\rho_c$)}
To Evaluate the Eq. \ref{FPSLEkr} we first consider the case $x_t^\infty \gg \zeta_t$ i. e. for short times of the evolution. In this limit $x_t^{\infty}\approx x^{\infty}$ and $\zeta_t \ll x^{\infty}$ so that we can treat $\frac{\zeta_t}{x^{\infty}}$ as the perturbation parameter. Expanding the second term in Eq \ref{FPSLEkr} in powers of this parameter we reach the formula:
\begin{equation}
\begin{split}
I&\equiv \langle \frac{1}{(\zeta_t - x_t^\infty)}\delta(x-x_t)\delta(y-y_t)\rangle \\
&=\sum_{n=0}^{\infty} \left\langle \frac{1}{x_t^\infty}\left(\frac{\zeta_t}{x_t^\infty}\right) ^n\delta(x-x_t)\delta(y-y_t)\right\rangle\\
& \simeq \sum_{n=0}^{\infty} \frac{1}{(x^\infty)^{n+1}} \left\langle \left(\zeta_t\right) ^n\delta(x-x_t)\delta(y-y_t)\right\rangle
\end{split}
\end{equation}
To the first order of $\frac{\zeta_t}{x^{\infty}}$ we have:
\begin{equation}
I\simeq \langle (\frac{1}{x_t^\infty} - \frac{\zeta_t}{(x_t^\infty)^2})\delta(x-x_t)\delta(y-y_t)\rangle.
\end{equation}
Taking $x_t^\infty\simeq x^\infty$ and using the equation $x_t+\zeta_t=-\int_{0}^{t}\frac{-2x_s}{x_s^2+y_s^2}\,ds$, we can rewrite the above equation as:
\begin{equation}
\begin{split}
I&\simeq \frac{-1}{x^\infty}\rho_c+\frac{1}{(x^\infty)^2}\langle (x_t + \int_{0}^{t}\frac{-2x_s}{x_s^2+y_s^2}\,ds)\delta(x-x_t)\delta(y-y_t)\rangle \\
&=\frac{-1}{x^\infty}\rho_c+\frac{1}{(x^\infty)^2}(x\rho_c+2\tilde{I})\\
&\tilde{I}\equiv \langle \int_{0}^{t}\frac{-2x_s}{x_s^2+y_s^2}\,ds\delta(x-x_t)\delta(y-y_t)\rangle.
\end{split}
\end{equation}
To proceed, we calculate $\tilde{I}$:
\begin{equation}
\begin{split}
\tilde{I}&= \langle \int_{0}^{t} ds \int \frac{x^{\prime}}{{x^{\prime}}^2+{y^{\prime}}^2}\delta(x^{\prime}-x_s)\delta(y^{\prime}-y_s)\delta(x-x_t)\delta(y-y_t)\rangle\ dx^{\prime}dy^{\prime}\\
&=\int_{0}^{t} ds \int \frac{x^{\prime}}{{x^{\prime}}^2+{y^{\prime}}^2}\rho_2(x,y,t;x^{\prime},y^{\prime},s)\ dx^{\prime}dy^{\prime}.
\end{split}
\end{equation}
In the above formula, $\rho_2(x,y,t;x^{\prime},y^{\prime},s)=\langle \delta(x^{\prime}-x_s)\delta(y^{\prime}-y_s)\delta(x-x_t)\delta(y-y_t)\rangle$ is the probability density of the tip of the curve being in $(x,y)$ at time $t$ and in $(x^{\prime},y^{\prime})$ at time $s$ ($s<t$). The result is:
\begin{equation}
\begin{split}
\frac{\partial\rho}{\partial t}&=\left[\frac{\kappa}{2}\partial_x^2 +\left(\frac{2x}{x^2+y^2}-\frac{\rho_c}{x^\infty}\right)\partial_x -\frac{2y}{x^2+y^2}\partial_y\right]\rho(x,y,t) \\
&+\frac{\rho_c}{(x^\infty)^2} \left[  \partial_x (x\rho(x,y,t)+ 2\int_{0}^{t} ds \int \frac{x^{\prime}}{{x^{\prime}}^2+{y^{\prime}}^2}\partial_x\rho_2(x,y,t;x^{\prime},y^{\prime},s)\ dx^{\prime}dy^{\prime} \right]   + O((x^\infty)^{-3}).
\end{split}
\label{FPESLEkr}
\end{equation}
The above formula can be generalized to any order of the perturbation parameter $\frac{\zeta_t}{x^{\infty}}$. The solutions of the above equation needs the information about $\rho_2$ and higher order probability functions which are less important than $\rho$ for early times. For larger times one also may use the hypothesis $\rho_2(x,y,t;x^{\prime},y^{\prime},s)\approx \rho(x,y,t)\rho(x^{\prime},y^{\prime},s)$. Let us write every thing up to $O\left( \left(x^{\infty} \right)^{-1}\right)$, i.e. $\rho(x,y,x^{\infty},t)=\rho(x,y,t)+\epsilon \rho_1(x,y,t)$ in which $\epsilon\equiv \frac{r}{x^{\infty}}$. Therefore, putting this function into Eq. \ref{FPESLEkr}, up to first order of $\epsilon$ we find that:
\begin{equation}
\frac{\partial\rho_1}{\partial t}=\left[\frac{\kappa}{2}\partial_x^2 +\frac{(\kappa+2)x}{x^2+y^2}\partial_x -\frac{2y}{x^2+y^2}\partial_y+\frac{1}{r^4}\left((\kappa-4)y^2+4x^2\right)\right]\rho_1(x,y,t) 
\end{equation}
In this case we also have the global scale transformation symmetry and translational invariance, resulting to the same form for $\rho_1$ as SEC \ref{FP}, i.e. $\rho_1=t^{-1}\psi(\chi,\eta)$ for $\psi$ being a function to be determined. Straightforward calculations yield:
\begin{equation}
\begin{split}
& a_{\chi\chi}\partial_{\chi}^2\psi+2a_{\chi\eta}\partial_{\chi}\partial_{\eta}\psi+a_{\eta\eta}\partial_{\eta}^2\psi+a_{\chi}\partial_{\chi}\psi+a_{\eta}\partial_{\eta}\psi+a\psi=0\\
& a_{\chi\chi}=2\kappa\frac{\eta^2\chi}{1+\eta^2}\\
& a_{\eta\eta}=\frac{\kappa}{2}\frac{(1+\eta^2)}{\chi}\\
& a_{\chi\eta}=\kappa\eta\\
& a_{\chi}=\kappa+\chi+\frac{2((\kappa+2)\eta^2-2)}{\eta^2+1}\\
& a_{\eta}=(\kappa+4)\frac{\eta}{\chi}\\
& a=1+\frac{1}{2\chi(1+\eta^2)}\left((\kappa+2)\eta^2-2\right)
\end{split}
\label{FPESLEkr-t-indep}
\end{equation}
 
which is parabolic PDE as the Eq \ref{FPE-t-indep}, since $a_{\chi\eta}^2-a_{\chi\chi}a_{\eta\eta}=0$ and its canonical form is ($\xi\equiv \frac{1+\eta^2}{\chi}$ as the previous sections):
\begin{equation}
\begin{split}
& \frac{1}{2}\kappa\xi\partial_{\eta}^2\psi-\left(\kappa+\frac{1+\eta^2}{\xi}+2\frac{(\kappa+2)\eta^2-2}{\eta^2+1}\right)\left(\frac{\xi^2}{1+\eta^2}\right)\partial_{\xi}\psi\\
& +(\kappa+4)\frac{\eta\xi}{1+\eta^2}\partial_{\eta}\psi+\left[ 1+\frac{\xi}{2(1+\eta^2)^2}\left( (\kappa+4)+4\eta^2\right)\right]\psi=0
\end{split}
\label{FPE3}
\end{equation}
If we repeat the same steps as SEC \ref{FP} we will find that $\Pi(\eta)\sim \eta^{-(8+\kappa)/\kappa}$, in which we have separated $\psi=\Pi_1(\eta)\Sigma_1(\xi)$ for large amounts of $\eta$. These results are gathered in the following equation:
\begin{equation}
\begin{split}
\rho(x,y,x^{\infty},t)&=\frac{\rho_0}{t}\Sigma(\xi)\left(\Pi(\eta)+\epsilon\Pi_1(\eta)\right)\\
& = \frac{\rho_0}{t}\Sigma(\xi)\left(\frac{a_0}{\eta^{(8-\kappa)\kappa}}+\frac{r}{x^{\infty}}\frac{a_1}{\eta^{(8+\kappa)/\kappa}}\right)
\end{split}
\end{equation}
in which $a_0$ and $a_1$ are some proportionality constants which can not be determined in this approach. The other way to obtain the behavior of $x_t$ and $y_t$ in terms of $t$ more exactly, is by using the Langevin equation Eq. \ref{BKWComponentsSLEkr}. Therefore in the next sub-section we analyze other properties of the growing SLE traces in dipolar (boundary to boundary) set up, first by its Langevin equations and secondly by means of CFT techniques. 
\subsection{Langeving equation for the long-time limit}
\begin{figure}
\centerline{\includegraphics[scale=.50]{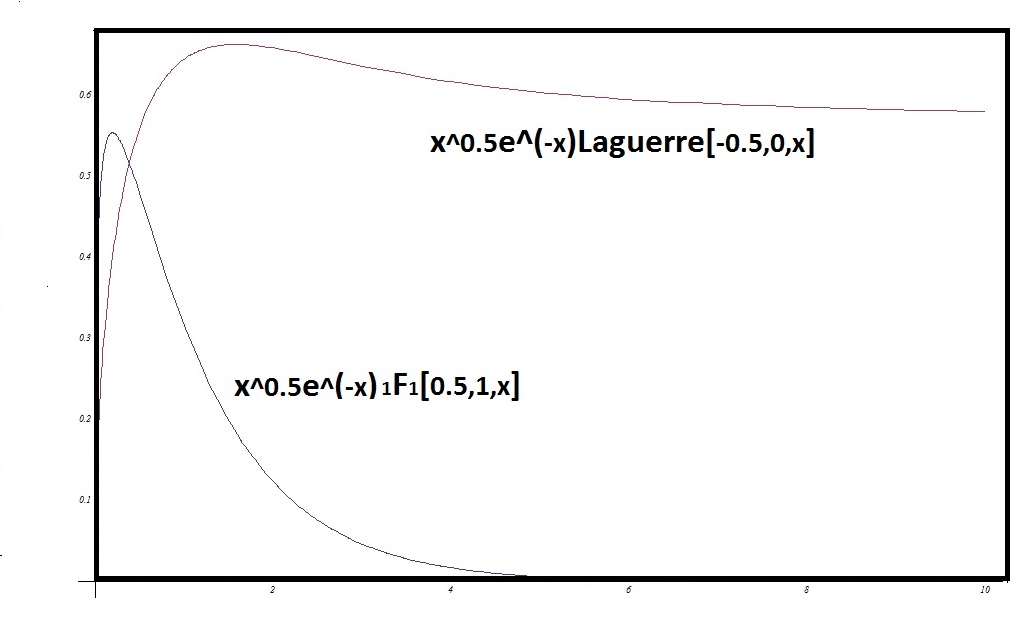}}
\caption{The behavior of the hypergeometric and Laguerre functions for $\kappa=4$ and $\rho_c=2$.}
\label{hyperLaguerre}
\end{figure}
The long-time limit of Eq \ref{BKWComponentsSLEkr} ($\frac{2x_t}{x_t^2+y_t^2}\simeq 0$ and $\zeta_t\gg x_t^{\infty}$) is \footnote{Throughout this subsection be use $\rho_c$ and assume that $\rho_c=\kappa-6$, although it is true for general $\rho_c$.}:
\begin{equation}
\begin{split}
& dx_t\simeq d\zeta_t\simeq -\sqrt{\kappa}dB_t - \frac{\rho_c}{\zeta_t}dt\\
& y_t\simeq\int_{-\infty}^{\infty}\frac{2y_t}{x^2+y_t^2}P_{\text{Bess}}(x,t)dx
\end{split}
\label{LargTimeBessel}
\end{equation}
in which $P_{\text{Bess}}(x,t)$ is the probability distribution of the Bessel process, i.e. the first expression of Eq \ref{LargTimeBessel}. In fact the Fokker-Planck equation for this Langevin equation is readily:
\begin{equation}
\partial_tP_{\text{Bess}}=-\rho_c\partial_x\left( \frac{P_{\text{Bess}}}{x}\right)+\frac{\kappa}{2}\partial_x^2P_{\text{Bess}}.
\label{FPBessel1}
\end{equation}
By changing variable $\zeta\equiv \frac{x^2}{2\kappa t}$ we obtain 
\begin{equation}
\zeta\frac{\text{d}^2}{\text{d}\zeta^2}P_{\text{Bess}}+\left( \frac{\kappa-2\rho_c}{2\kappa}+\zeta\right)\frac{\text{d}}{\text{d}\zeta}P_{\text{Bess}}+\frac{\rho_c}{2\kappa\zeta}P_{\text{Bess}}=0.
\label{FPBessel2}
\end{equation}
By the transformation $P_{\text{Bess}}=\zeta^{\frac{1}{2}}e^{-\zeta}w(\zeta)$ we see that $w(\zeta)$ satisfies the Kummer's equation:
\begin{equation}
\zeta \frac{\text{d}^2}{\text{d}\zeta^2}w+\left(\frac{3\kappa-2\rho_c}{2\kappa}-\zeta\right)\frac{\text{d}}{\text{d}\zeta}w-\left( 1-\frac{\rho_c}{\kappa}\right)w=0 
 \end{equation}
whose solution is:
\begin{equation}
w(\zeta)=A _1\text{F}_1\left[ 1-\frac{\rho_c}{\kappa},\frac{3}{2}-\frac{\rho_c}{\kappa},\zeta\right]+B \text{L}^{\frac{1}{2}-\frac{\rho_c}{\kappa}}_{\frac{\rho_c}{\kappa}-1}\left[\zeta\right]
 \end{equation}
in which the first term is the hypergeometric function and the second term is the Laguerre polynomial. The second independent solution could have been assumed to be $\zeta^{-\frac{\kappa-2\rho_c}{2\kappa}} {_1}\text{F}_1\left[ \frac{1}{2},\frac{\kappa+2\rho_c}{2\kappa},\zeta\right]$, but the Laguerre function is more suitable as becomes clear in the following. In FIG \ref{hyperLaguerre} we have presented the plot of the two functions stated above. We see that as $\frac{x}{2\kappa t}=\zeta\rightarrow \infty$, the hypergeometric function goes to zero, in contrast to the Laguerre function which remains finite. So we conclude that $B=0$. Now let us calculate $\left\langle x\right\rangle $ and $\left\langle \frac{2y_t}{x^2+y_t^2}\right\rangle$ which are:
\begin{equation}
\begin{split}
& \left\langle x\right\rangle=\sqrt{2\kappa t}\frac{ \int_{0}^{\infty}\sqrt{\zeta}e^{-\zeta}{_1}\text{F}_1\left[ 1-\frac{\rho_c}{\kappa},\frac{3}{2}-\frac{\rho_c}{\kappa},\zeta\right]d\zeta}{\int_{0}^{\infty}e^{-\zeta}{_1}\text{F}_1\left[ 1-\frac{\rho_c}{\kappa},\frac{3}{2}-\frac{\rho_c}{\kappa},\zeta\right]d\zeta}\\
&\left\langle \frac{2y_t}{x^2+y_t^2}\right\rangle=\frac{ \int_{0}^{\infty}\frac{2y_t}{2\kappa t \zeta+y_t^2}e^{-\zeta}{_1}\text{F}_1\left[ 1-\frac{\rho_c}{\kappa},\frac{3}{2}-\frac{\rho_c}{\kappa},\zeta\right]d\zeta}{\int_{0}^{\infty}e^{-\zeta}{_1}\text{F}_1\left[ 1-\frac{\rho_c}{\kappa},\frac{3}{2}-\frac{\rho_c}{\kappa},\zeta\right]d\zeta}
\end{split}
\label{average}
\end{equation}
The solution of the first expression of the above formula $\left\langle x\right\rangle$ is:
\begin{equation}
\begin{split}
& \left\langle x\right\rangle=f\sqrt{2\kappa t} \\
& f=\frac{\pi^{\frac{3}{2}}\sec(\pi \frac{\rho_c}{\kappa})}{4\Gamma(\frac{1}{2}+\frac{\rho_c}{\kappa})}\left\lbrace \frac{2{_2}F_1(\frac{1}{2},1-\frac{\rho_c}{\kappa},\frac{1}{2}-\frac{\rho_c}{\kappa},1)}{\Gamma(\frac{1}{2}-\frac{\rho_c}{\kappa})}-\frac{\Gamma(1+\frac{\rho_c}{\kappa})}{\Gamma(1-\frac{\rho_c}{\kappa})}\frac{{_2}F_1(\frac{3}{2},1+\frac{\rho_c}{\kappa},\frac{3}{2}+\frac{\rho_c}{\kappa},1)}{\Gamma(\frac{3}{2}+\frac{\rho_c}{\kappa})}\right\rbrace
\end{split}
\label{average_x}
\end{equation}
which shows that in the long-time limit we have $\left\langle x\right\rangle\sim t^{\frac{1}{2}}$ as the chordal case and the presence of the preferred point only changes the proportionality constant of this dependence. To obtain the amount of $y_t$, we note that the answer of the second equation in Eq \ref{average} is:
\begin{equation}
\int_{0}^{\infty}\frac{1}{a\zeta+b^2}e^{-\zeta}{_1}\text{F}_1\left[ 1-\frac{\rho_c}{\kappa},\frac{3}{2}-\frac{\rho_c}{\kappa},\zeta\right]d\zeta =\frac{1}{a} G^{31}_{23}\left(^{0,\frac{1}{2}}_{0,0,-\frac{1}{2}+\frac{\rho_c}{\kappa}}|\frac{b^2}{a}\right)
\end{equation}
in which $G^{mn}_{pq}\left(^{a_1,a_2,...,a_p}_{b_1,b_2,...,b_q}|z\right)$ is the Meijer G-function \cite{Meijer}. We have ($\chi_t\equiv \frac{y_t}{\sqrt{2\kappa t}}$ as above and $w\equiv \ln t$):
\begin{equation}
\frac{1}{\chi_t}\frac{\text{d}\chi_t}{\text{d}w}=-\frac{1}{2}+\frac{\sqrt{\pi}}{2\kappa \Gamma(\frac{1}{2}+\frac{\rho_c}{\kappa})}G^{31}_{23}\left(^{0,\frac{1}{2}}_{0,0,-\frac{1}{2}+\frac{\rho_c}{\kappa}}|\chi_t^2\right)
\end{equation}
It is difficult to find the general solution of the above differential equation, but if we expand
\begin{equation} 
\frac{1}{\chi_t\left(-\frac{1}{2}+ \frac{\sqrt{\pi}}{2\kappa \Gamma(\frac{1}{2}+\frac{\rho_c}{\kappa})}G^{31}_{23}\left(^{0,\frac{1}{2}}_{0,0,-\frac{1}{2}+\frac{\rho_c}{\kappa}}|\chi_t^2\right)\right) }=\alpha_0+\alpha_1(\chi_t-1)+O((\chi_t-1)^2)
\end{equation}
we will be able to determine the above integral as follows:
\begin{equation}
\chi_t=\frac{y_t}{\sqrt{2\kappa t}}\approx 1+\frac{\alpha_0}{\alpha_1}\left( -1+\left( 1+2\frac{\alpha_1}{\alpha_0^2}\ln t\right)^{\frac{1}{2}}\right)
\end{equation}
which shows the logarithmic deviation from the ordinary SLE solution, i.e. $y_t^{\text{Chordal}}\sim \sqrt{2\kappa t}$. $\alpha_0$ and $\alpha_1$ in the above formula are:
\begin{equation}
\begin{split}
& \alpha_0=\frac{1}{-\frac{1}{2}+\frac{\sqrt{\pi}}{2\kappa \Gamma(\frac{1}{2}+\frac{\rho_c}{\kappa})}G^{31}_{23}\left(^{0,\frac{1}{2}}_{0,0,-\frac{1}{2}+\frac{\rho_c}{\kappa}}|1\right) }\\
& \alpha_1=\frac{2+8\frac{\sqrt{\pi}}{2\kappa \Gamma(\frac{1}{2}+\frac{\rho_c}{\kappa})}G^{31}_{23}\left(^{-1,-\frac{1}{2}}_{-1,0,-\frac{3}{2}+\frac{\rho_c}{\kappa}}|1\right)-4\frac{\sqrt{\pi}}{2\kappa \Gamma(\frac{1}{2}+\frac{\rho_c}{\kappa})}G^{31}_{23}\left(^{0,\frac{1}{2}}_{0,0,-\frac{1}{2}+\frac{\rho_c}{\kappa}}|1\right)}{\left(1-2\frac{\sqrt{\pi}}{2\kappa \Gamma(\frac{1}{2}+\frac{\rho_c}{\kappa})}G^{31}_{23}\left(^{0,\frac{1}{2}}_{0,0,-\frac{1}{2}+\frac{\rho_c}{\kappa}}|1\right)\right)^2}.
\end{split}
\end{equation}
Therefore we see that the vertical component ($y_t$) of the tip of the SLE trace shows serious deviations from the ordinary SLE results, in contrast to the horizontal component which has the same dependence (with modified proportionality constant):
\begin{equation}
\left\lbrace \begin{array}{cc} x_t= f\sqrt{2\kappa t} \\  y_t\simeq \sqrt{2\kappa t}+\frac{\alpha_0}{\alpha_1}\sqrt{2\kappa t}\left[\left( 1+2\frac{\alpha_1}{\alpha_0^2}\ln t\right)^{\frac{1}{2}}-1\right] \end{array}\right.
\end{equation}
\subsection{CFT Prediction for the PDF of SLE($\kappa,\rho_c$) Traces}
As SEC \ref{CFTchordal}, in this section we apply the CFT constraints to the correlation function which are representative of the probability density of the SLE($\kappa,\rho_c$) random curves. For the this case we can use the Eqs \ref{LPE} and \ref{Q-P} to obtain 
\begin{equation}
\begin{split}
[& \frac{\kappa}{4}\partial_{\zeta_0}^2- 2h_F\textrm{Re}\left( \frac{1}{(z-\zeta_0)^2} \right)+\frac{1}{z-\zeta_0} \frac{\partial}{\partial z}+\frac{1}{\bar{z}-\zeta_0} \frac{\partial}{\partial \bar{z}}\\
&+\frac{1}{x^\infty-\zeta_0} \left((6-\kappa)\frac{\partial}{\partial \zeta_0}+2\frac{\partial}{\partial x^\infty}\right) ]P(x,y,\zeta_0,x^{\infty})=0.
\end{split}
\label{LPE_SLEkr}
\end{equation}
After some straightforward calculations, using the fact that $P=\frac{1}{y^{2h_F}}h(u)$ (Eq. \ref{FPF}) we obtain:
\begin{equation}
\left[ \frac{\kappa}{2}\partial_{\zeta_0}^2+\frac{2(x-\zeta_0)}{(x-\zeta_0)^2+y^2}\partial_x-\frac{2y}{(x-\zeta_0)^2+y^2}\partial_y+\frac{6-\kappa}{x^{\infty}-\zeta_0}\partial_{\zeta_0}+\frac{2}{x_{\infty}-\zeta_0}\partial_{x^{\infty}}+8h_F\frac{y^2}{\left( x^2+y^2\right)^2}\right]h(u)=0
\end{equation}
Rearranging the derivatives in terms of $u=\text{Re}\frac{\left(z-\zeta_0\right)\left(\bar{z}-x^{\infty}\right)}{y\left(x^{\infty}-\zeta_0\right)}=\frac{(x-\zeta_0)(x-x^{\infty})}{y(x^{\infty}-\zeta_0)}+\frac{y}{x^{\infty}-\zeta_0}$, after some algebra it is concluded that:
\begin{equation}
\begin{split}
\left[\frac{\kappa}{2}\left(1+u^2\right)^2\partial_{u}^2+4u\left(1+u^2\right)\partial_{u}+8h_F\right]h(u)=0
\end{split}
\end{equation}
which is the same as Eq \ref{eta_Eq} by replacing $\eta\rightarrow u$. Therefore for the case SLE($\kappa,\rho_c=\kappa-6$) we have $P(x,y,\zeta_0,x^{\infty},\epsilon)=\left(\frac{\epsilon}{y}\right)^{2h_F}h(u)=\left(\frac{\epsilon}{y}\right)^{2-d_f}\left(1+u^2\right)^{-(8-\kappa)/(2\kappa)}$.\\
This result is expected, since due to conformal invariance of the model we can generate SLE( $\kappa,\kappa-6$) curves from chordal ones by the map $z=\phi(w)=\frac{x^{\infty}w}{x^{\infty}-w}$ in which $z$ indicates the coordinate in which the chordal SLE grows and $w=x+iy$ the dipolar one. Therefore we expect that ($\epsilon^{\prime}$ is the transformed radius under $\phi$)
\begin{equation} P_{\text{dipolar}}(w,x^{\infty})=P_{\text{chordal}}(\phi(w))=\left(\frac{\epsilon^{\prime}}{\text{Im}[\phi(w)]}\right)^{2-d_f}h(\eta^{\prime}=\frac{\text{Re}[\phi(w)]}{\text{Im}[\phi(w)]})
\end{equation}
We can easily verify that $\frac{\text{Re}[\phi(w)]}{\text{Im}[\phi(w)]}=-u$ and since $h(u)$ is even with respect to $u$ we find that $P_{\text{dipolar}}(w,x^{\infty})=\left(\frac{\epsilon^{\prime\prime}}{y}\right)^{2-d_f}h(u)$ in which $\epsilon^{\prime\prime}\equiv \frac{\left((x-x^{\infty})^2+y^2\right)}{{x^{\infty}}^2}\epsilon^{\prime}$ is the renormalized radius.
\section{Conclusion}
In this paper, we analyzed the statistics of the SLE curves described by SLE($\kappa,\rho_c$). We obtained the Fokker-Planck equation governing these SLE traces. We found that the long-time limit behavior of the $x$ and $y$ components of the tip of the growing curves are not the same. The $x$ component is only renormalized (meaning that the proportionality constant changes), and the $y$ component is changed considerably, i.e. its functional form changes. The exact forms of these changes were reported in this paper.

\end{document}